\documentclass{article}

\usepackage{arxiv}

\usepackage[utf8]{inputenc} 
\usepackage[T1]{fontenc}    
\usepackage{hyperref}       
\usepackage{url}            
\usepackage{booktabs}       
\usepackage{amsfonts}       
\usepackage{nicefrac}       
\usepackage{microtype}      
\usepackage{lipsum}		
\usepackage{graphicx}
\usepackage{natbib}
\usepackage{doi}

\usepackage{diagbox}

\usepackage{color}
\usepackage[normalem]{ulem} 
\usepackage{amsmath,amssymb,enumerate,xspace}

\title{The Bottleneck Birthday Problem}

\author{{Chijul B.~Tripathy}\\
	Homestead High School\\
	Cupertino, CA 95014 \\
	\texttt{chijul.b.tripathy@gmail.com} \\
}

\date{}

\renewcommand{\shorttitle}{}

\hypersetup{
pdftitle={The Bottleneck Birthday Problem},
pdfauthor={Chijul B.~Tripathy},
pdfkeywords={The birthday problem, Bottleneck birthday problem, Stirling number of the second kind, restricted Stirling number of the second kind, Dynamic programming, Recurrence},
}

\newcommand{\gobbleOut}[1]{} 

\definecolor{dukeblue}{rgb}{0.0, 0.0, 0.61}
\definecolor{harvardcrimson}{rgb}{0.79, 0.0, 0.09}
\definecolor{dartmouthgreen}{rgb}{0.05, 0.5, 0.06}
\definecolor{gold(metallic)}{rgb}{0.83, 0.69, 0.22}
\definecolor{goldenrod}{rgb}{0.85, 0.65, 0.13}
\definecolor{indianyellow}{rgb}{0.89, 0.66, 0.34}
\definecolor{indianred}{rgb}{0.8, 0.36, 0.36}
\definecolor{eggplant}{rgb}{0.38, 0.25, 0.32}

\newcommand{\bbp}{\texttt{BBP}\xspace}

\newcommand{\set}[1]{\ensuremath{\left\{#1\right\}}}
\newcommand{\cset}[2]{\ensuremath{\left\{\!\!\begin{array}{l}#1\end{array}\left|\begin{array}{l}#2\end{array}\right.\!\!\right\}}}

\newcommand{\paren}[1]{\ensuremath{\left(#1\right)}}
\newcommand{\bracket}[1]{\ensuremath{\left[#1\right]}}
\newcommand{\ang}[1]{\ensuremath{\left\langle#1\right\rangle}}

\newcommand{\prob}[1]{\ensuremath{P\!\left(#1\right)}}

\newcommand{\ceil}[1]{\ensuremath{\left\lceil\!\!\begin{array}{l}#1\end{array}\!\!\right\rceil}}

\newcommand{\ProbabilityStuff}[2]{\ensuremath{\mathop{\textup{#1}}\nolimits\left[#2\right]}}
\newcommand{\Exp}[1]{\ProbabilityStuff{E}{#1}}

\newcommand\card[1]{\ensuremath{\left|#1\right|}}

\newcommand{\shorthand}[1]{{#1}\xspace}

\newcommand{\eg}{\shorthand{e.g.}}

\newcommand{\BigOh}[1]{\ensuremath{\operatorname{O}\!\left(#1\right)}}

\newcommand{\maxn}{\ensuremath{n_{\max}}\xspace}

\begin{document}
\maketitle

\begin{abstract}
	We introduce a fun problem that can be considered as a variant of the classic birthday problem, the \emph{Bottleneck Birthday Problem (\bbp)}. It is stated as: what is the maximum number of people we have to choose so that no day of the year has more than $r \geq 1$ birthdays incident on it with probability at least 1/2? We provide a survey of techniques used in the literature on occupancy and load balancing problems to derive recurrence relations for exact computation of the probability, and the number of people keeping probability fixed at a threshold. Further, we show that restricted Stirling numbers of the second kind can be used to derive an additional recurrence, in a novel way. We provide complexity comparisons and numerical results from an implementation of the recurrences.
\end{abstract}


\section{Introduction}\label{sec:introduction}

What is the probability that from a group of $n$ randomly chosen people, at least two individuals will have the same birthday? This question is famously known as the \emph{birthday problem} (see~\cite{Feller1968introductionV1}, page 33), having been around for nearly 100 years. Austrian mathematician Richard von Mises published this in 1939~\cite{vonMises1939}. However, English mathematician Harold Davenport is typically credited for this problem in 1927, though he did not publish it due to its simplicity and counterintuitive nature.

A standard set up for the birthday problem assumes that birthdays are uniformly distributed over the days of the year. Further, it relies on a few simplifying assumptions including but not limited to no selection bias (\eg, no specific preference for certain birthdays or months), no interaction between individuals’ birthdays (\eg, no external factors, such as hospital scheduling affecting birth times), and no multiple births such as twins, among others. Under these assumptions, it can be shown that only 23 people are needed to achieve a probability of more than 1/2 that at least two people share the same birthday. The result seems counterintuitive, since by the Pigeonhole principle, we would need 366 people to guarantee a shared birthday, considering 365 days being a standard year. 

In the past several decades, many variants and extensions to the classic birthday problem have been published in the literature~\cite{Naus1968,mosteller1962,Mathis1991,Nunnikhoven1992,DasGupta2005,inoue2008}. For example, an interesting variation is the \emph{Almost Birthday Problem} which asks for the minimum number of people needed so that at least two have birthdays within $d$ consecutive days~\cite{Naus1968}. For $d = 1, 2, 3, 4, 5$, the corresponding minimum number of people needed are $n = 23, 14, 11, 9, 8$, respectively. Many of these variations are intriguing puzzles in their own right. Computationally efficient approximation algorithms to compute the solutions have been an area of great interest in the research community, since several real-world problems in the domains of computer science including hashing, cryptography~\cite{bellare2004}, communication networks and load balancing in distributed systems~\cite{eijkhout2019}, and problems in other domains including biology~\cite{Green2024} and medicine~\cite{Krawitz2015}, forensic science~\cite{Kaye2013Beyond}, statistics, social science and data science~\cite{gold2024}, can be modeled as one of the variants of the birthday problem. In a recent work~\cite{tripathy2025}, we derived recurrence relations and designed exact algorithms for a variant of the the birthday problem called the \emph{Strong Birthday Problem}, first introduced in~\cite{DasGupta2005}.

In this article, we consider an extension of the birthday problem, which we name as the \emph{Bottleneck birthday problem} (\bbp). The problem asks to find the maximum number of people we have to choose so that no day of the year has more than $r \geq 1$ birthdays incident on it with probability at least 1/2. In the context of hashing, this problem previously appeared in the literature~\cite{Ramakrishna1987}, who gave a recurrence relation to compute the probability. We survey this work, and derive multiple recurrence formulas to solve the problem. We show a novel connection to the restricted Stirling numbers of the second kind, which we used to formulate the recurrence. We also give numerical results based on our implementations of the algorithms to solve \bbp, discuss applications of this problem, and provide a note on the future directions that we think may lead to interesting research.

\section{The Bottleneck Birthday Problem (\bbp)}\label{sec:bottleneckbirthdayproblem}
\noindent{\bf Problem Definition (Bottlenect Birthday Problem): }  Let $m \geq 1$ be the number of days in a year, and let $n \geq 1$ be the number of people whose birthdays are independently and uniformly distributed over the $m$ days. That is, the probability that $i$th person's birthday is incident on $j$th day of the year is $1/m$, for $1 \leq i \leq n$ and $1 \leq j \leq m$. This is a simplifying assumption for the problem, but it is a reasonable one in practical settings and may allow for extension to include other distributional assumptions. Let $r \geq 1$ be an integer threshold and $\gamma \in \bracket{0, 1}$ be a real number denoting the probability threshold.

In this setting, we can define \bbp as follows: what is the maximum number \maxn of people needed such that no day has more than $r$ birthdays incident on it with probability at least $\gamma$. Formally,
\begin{eqnarray}
    \maxn = \max\cset{n}{\paren{\sum_{i = 1}^m X_i = n} \land \paren{\prob{\displaystyle\max_{i=1}^m \{X_i\} \leq r} \geq \gamma}},
\end{eqnarray}
where $X_i$ denotes the number of people with birthdays incident on day $i$, for $1 \leq i \leq m$. 

Since the joint distribution of $X_1, \ldots, X_m$ follows a multinomial distribution, the probability that $n$ people's birthdays are distributed over $m$ days so that no day has more than $r$ birthdays incident on it can be written as
\begin{eqnarray}
    \prob{\max\{X_1, \ldots, X_m\} \leq \gamma} &=& \frac{1}{m^n} \sum_{\substack{k_1, \ldots, k_m \geq 0 \\ k_1 + \cdots + k_m = n \\ \max\{k_1, \ldots, k_m\} \leq r}} \binom{n}{k_1, k_2, \ldots, k_m},
\end{eqnarray}
where $\binom{n}{k_1, \ldots, k_m} = \frac{n!}{k_1! k_2! \cdots k_m!}$ is the multinomial coefficient. The denominator $m^n$ is the size of the sample space.

The computational complexity of evaluating the above formula is exponential since we have to enumerate $\BigOh{\binom{n + m - 1}{m - 1}} \approx \frac{(n + m - 1)^m} {m!}$ configurations.

We note that \bbp can be thought of as a complementary problem to the classic birthday problem.

\section{Recurrences for \bbp}
In this section, we give several recurrences, that appeared in the literature and the ones we have come up with. These recurrences compute the probability $\prob{m, n, r}$ for the input values $m$, $n$ and $r$. For a given value of the probability threshold $\gamma$, say, $\gamma = 1/2$, binary search on the values of $n$ can be performed to find \maxn. When we note the space and time complexities of the algorithms, we only note the combinatorial complexity and discard the bit-complexity of mathematical operations involving large numbers.

\subsection{A Recurrence for \bbp Considering One Day at a Time}\label{sec:recOneDay}

The probability that exactly $k \leq r$ birthdays are incident on day $j$ is $\binom{n}{k} \paren{\frac{1}{m}}^k \paren{1 - \frac{1}{m}}^{n - k}$, and the remaining $n - k$ birthdays are distributed among the remaining $m - 1$ days is $\prob{m - 1, n - k, r}$. By the law of total probability, summing over $k = 0, \ldots, \min(n, r)$, we obtain
\begin{eqnarray}
   \prob{m, n, r} &=& \sum_{k = 0}^{\min(n, r)} \binom{n}{k} \paren{\frac{1}{m}}^k \paren{1 - \frac{1}{m}}^{n - k} \prob{m - 1, n - k, r}.\label{eqn:odtrecurrence}
\end{eqnarray}

The base cases are:
\begin{align*}
\prob{1, n, r} &= \begin{cases} 1, & \text{if } n \leq r; \\ 0, & \text{if } n > r. \end{cases} \\
\prob{m, 0, r} &= 1, \quad \text{for all } m \geq 1. \\
\prob{0, n, r} &= 0, \quad \text{for all } n \geq 1.
\end{align*}

The above recurrence can be implemented using dynamic programming. The space complexity is \BigOh{mn} and the time complexity is \BigOh{mnr}. A space-efficient rolling array implementation has \BigOh{n} space complexity.

We note that the above recurrence is a direct application of the law of total probability to balls-and-bins problems derived from the first principles. Similar analysis for the load balancing problems appear in~\cite{kolchin1978,Raab1999,Azar1999} for various settings.

\subsection{A Recurrence for \bbp Based on Counting}\label{sec:countingrecurrence}
We derive a recurrence purely based on counting the number of ways we can arrange the birthdays in the setting of the problem \bbp. Let $T(m, n, k, r)$ be the number of ways in which $n$ people's birthdays are distributed over $k$ days out of $m$ possible days in a year so that no day has more than $r$ birthdays incident on it. Clearly, $n \leq kr$. To set up a recurrence for $T(m, n, k, r)$, let's consider the following two mutually exclusive cases when we add the $n$th person.

\begin{enumerate}[{\bf {Case} 1:}]
    \item $n$th person forms a singleton birthday. There are $T(m, n - 1, k - 1, r)$ ways in which the first $n - 1$ people's birthdays can be distributed over $k - 1$ days so that each day has at most $r$ people's birthdays incident on it. The singleton birthday can be one of the remaining $m - (k - 1) = m - k + 1$ days. Therefore, the number of ways this can be done is $(m - k + 1) T(m, n - 1, k - 1, r)$.

    \item $n$th person's birthday is incident on one of the $k$ days containing $n - 1$ birthdays. Since there are $T(m, n - 1, k, r)$ ways in which the first $n - 1$ people's birthdays are distributed over $k$ days so that each day has at most $r$ people's birthdays incident on it, the addition of $n$th person can be done in $k T(m, n - 1, k, r)$ ways. However, we must subtract the cases where $n$th person's addition leads to the cases when the number of birthdays incident on a particular day goes from $r$ to $r + 1$. The number of ways in which a day with $r$ birthdays incident on it, to which $n$th person's birthday will be added, can be formed in all of $T(m, n - 1, k, r)$, is by choosing $r$ people from $n - 1$ people in $n - 1 \choose r$ ways and assigning them one of the $m$ days as birthday, and then partitioning the remaining $n - 1 - r$ people into $k - 1$ days as birthdays out of $m - 1$ possible days, where each day has at most $r$ people's birthdays incident on it. This can be done in $m {n - 1 \choose r} T(m - 1, n - 1 - r, k - 1, r)$ ways.
\end{enumerate}

Combining the above two cases we write the recurrence relation as
\begin{eqnarray}
     T(m, n, k, r) &=& (m - k + 1) T(m, n - 1, k - 1, r) + k T(m, n - 1, k, r) \nonumber \\
     && - m {n - 1 \choose r} T(m - 1, n - 1 - r, k - 1, r).
\end{eqnarray}

The base cases are $T(m, 0, 0, r) = 1$, and $T(m, n, k, r) = 0$ when $m \leq 0$ or $n \leq 0$ or $k \leq 0$ or $m - k + 1 \leq 0$ or $n > k r$ .

Therefore, the complete recurrence can be written as
\begin{eqnarray}
    T(m, n, k, r) &=& 
    \begin{cases}
        0, & \text{if } (n \leq 0 \land k > 0) \lor (n > 0 \land k \leq 0) \\ & \quad \lor (n < k) \lor (n > kr) \lor (m - k + 1 \leq 0); \\
        1, & \text{if $(n = 0 \land k = 0)$}; \\
        (m - k + 1) T(m, n - 1, k - 1, r) \\ + k T(m, n - 1, k, r) \\ - m {n - 1 \choose r} T(m - 1, n - 1 - r, k - 1, r), & \text{otherwise.}\label{eqn:countingrecurrence}
    \end{cases}
\end{eqnarray}

The above recurrence can be computed using dynamic programming.

For the bottleneck birthday problem, $\ceil{\frac{n}{r}} \leq k \leq \min(m, n)$. Therefore, the number of ways in which $n$ people can have birthdays out of $m$ possible days so that no day has more than $r$ people's birthdays incident on it is given by 
\begin{eqnarray}
    N &=& \sum_{k = \ceil{\frac{n}{r}}}^{\min(m, n)} T(m, n, k, r),
\end{eqnarray}
and hence the probability of this occurring is
\begin{eqnarray}
    \prob{m, n, r} &=& \frac{1}{m^n} N = \frac{1}{m^n} \sum_{k = \ceil{\frac{n}{r}}}^{\min(m, n)} T(m, n, k, r).
\end{eqnarray}

The above formula can be computed using dynamic programming. The time complexity of the dynamic programming implementation is \BigOh{m^2 n^2 r}. The space complexity is \BigOh{m n^2} and with optimization it can be reduced to \BigOh{m n r}. Clearly, this algorithm is a slow one and needs more space. However, when we need the actual count for specific $k$ values or do distribution analysis across different $k$ values, and understanding the structure of the solutions, this recurrence gives further insights.

Next,  we introduce the $r$-restricted Stirling number of the second kind, and use that to derive a recurrence to compute \prob{m, n, r}.

\subsection{Restricted Stirling Numbers of the Second Kind}\label{sec:RestrictedStirling2ndKind}
Stirling numbers of the \emph{second kind}, denoted by ${n \brace k}$, counts the number of ways to partition $n$ distinct objects \set{a_1, \ldots, a_n} into $k$ nonempty \emph{partitions (subsets).} For example, ${4 \brace 2} = 7.$

We can write a recurrence for ${n \brace k}$ as follows (see~\cite{graham94} page 259):

\begin{eqnarray}
    {n \brace k} &=& 
    \begin{cases}
        0, & \text{if $(n = 0 \land k > 0) \lor (n > 0 \land k = 0)$}; \\
        1, & \text{if $(n = 0 \land k = 0) \lor (n > 0 \land k = 1)$}; \\
        {n - 1 \brace k - 1} + k {n - 1 \brace k}, & \text{otherwise.}
    \end{cases}
\end{eqnarray}

The $r$-restricted Stirling number of the second kind, which we denote here by ${n \brace k}_{\leq r}$ is the number of ways of partitioning $n$ distinct objects, $a_1, \ldots, a_n$, into $\ceil{\frac{n}{r}} \leq k \leq n$ \emph{unlabeled} partitions where each partition has size at at most $r \geq 1$. Note that this is a restriction over the Stirling number of the second kind since no partition's size can exceed $r$.

For example, there ae 7 ways of partitioning the 4-element set \set{a_1, 
\ldots, a_4} into 2 partitions such that each partition has at most 3 elements:
\begin{eqnarray*}
    && \set{a_1, a_2, a_3}, \set{a_4} \quad \set{a_1, a_2, a_4}, \set{a_3} \quad \set{a_2, a_3, a_4}, \set{a_1} \quad \set{a_1, a_3, a_4}, \set{a_2} \\
    && \set{a_1, a_2}, \set{a_3, a_4} \quad \set{a_1, a_3}, \set{a_2, a_4} \quad \set{a_1, a_4}, \set{a_2, a_3}.
\end{eqnarray*}
Therefore, ${4 \brace 2}_{\leq 3} = 7$. Similarly, we can easily show that ${5 \brace 2}_{\leq 3} = 10$ and ${5 \brace 3}_{\leq 3} = 25$.

In order to formulate a recurrence, let's consider the $n$th object $a_n$. There are two mutually exclusive cases to consider.
\begin{enumerate}[{\bf {Case} 1:}]
    \item (Forms a partition). $a_n$ forms a singleton partition. There are ${n - 1 \brace k - 1}_{\leq r}$ ways in which the first $n - 1$ objects are partitioned into $k - 1$ partitions so that each partition has size $\leq r$. The singleton partition formed by $a_n$ can be added to each of the above ways of partitioning to form $k$ partitions of $n$ objects with each partition has size $\leq r$. Therefore, the number of ways this can be done is ${n - 1 \brace k - 1}_{\leq r}$.
    
    \item (Does not form a partition). $a_n$ is inserted into any of the $k$ nonempty partitions of the first $n - 1$ objects. Since the first $n - 1$ objects can be partitioned in ${n - 1 \brace k}_{\leq r}$ ways into $k$ nonempty partitions so that each partition has size at most $r$, addition of $a_n$ can be done in $k {n - 1 \brace k}_{\leq r}$ ways. However, we must subtract the cases where $a_n$'s addition leads to the partition size going from $r$ to $r + 1$. The number of ways in which $r$-size partitions can be formed in all of ${n - 1 \brace k}_{\leq r}$ partitions is by choosing $r$ objects from $n - 1$ objects in $n - 1 \choose r$ ways, and then partitioning the remaining $n - 1 - r$ objects into $k - 1$ partitions of size at most $r$, which in total is ${n - 1 \choose r} {n - 1 - r \brace k - 1}_{\leq r}$.
\end{enumerate}

Combining the above, we get the following recurrence for ${n \brace k}_{\leq r}$.
\begin{eqnarray}
    {n \brace k}_{\leq r} &=& {n - 1 \brace k - 1}_{\leq r} + k {n - 1 \brace k}_{\leq r} - {n - 1 \choose r} {n - 1 - r \brace k - 1}_{\leq r}.
\end{eqnarray}
The base cases are ${0 \brace 0}_{\leq r} = 1$, and ${n \brace k} = 0$ when $n \leq 0$ or $k \leq 0$ or $n > k r$. Also, ${n \brace n}_{\leq r} = 1$.

Therefore, the complete recurrence can be written as
\begin{eqnarray}
    {n \brace k}_{\leq r} &=& 
    \begin{cases}
        0, & \text{if $(n \leq 0 \land k > 0) \lor (n > 0 \land k \leq 0) \lor (n > kr)$}; \\
        1, & \text{if $(n = 0 \land k = 0)$}; \\
        {n - 1 \brace k - 1}_{\leq r} + k {n - 1 \brace k}_{\leq r} - {n - 1 \choose r} {n - 1 - r \brace k - 1}_{\leq r}, & \text{otherwise.}
    \end{cases}
\end{eqnarray}

See~\cite{Comtet1974,Howard1980,Zhao2008,Komatsu2015} for more on the restricted Stirling numbers of the second kind and related identities. We will use the above-derived recurrence for the restricted Stirling numbers of the second kind to formulate a recurrence for \bbp.

\subsection{A Recurrence for \bbp using the Restricted Stirling Numbers of the Second Kind}\label{sec:restrstr}
We can derive a recurrence using restricted Stirling numbers of the second kind as follows. The number of ways of distributing $n$ people's birthdays over $m$ possible days in a year so that each day has at most $r$ birthdays incident on it can be done as follows. Let's choosing $k$ birthdays out of $m$ days in ${m \choose k}$ ways for $\ceil{\frac{n}{r}} \leq k \leq \min(m, n)$. For each such choice, there are $k! {n \brace k}_{\leq r}$ ways of distributing $n$ people into $k$ days so that each day has at most $r$ people's birthdays incident on it. Here ${n \brace k}_{\leq r}$ is the $r$-restricted Stirling number of the second kind. Therefore, the total number of ways is given by
\begin{eqnarray}
    N &=& \sum_{k = \ceil{\frac{n}{r}}}^{\min(m, n)} {m \choose k} k! {n \brace k}_{\leq r},
\end{eqnarray}

Since the number of ways in which $n$ people can be assigned $m$ days as birthdays is $m^n$, which is the size of the sample space, the probability that $n$ people's birthdays are distributed over $m$ possible days in a year so that each day has at most $r$ birthdays incident on it is given by
\begin{eqnarray}
    \prob{m, n, r} &=& \frac{1}{m^n} N = \frac{1}{m^n} \sum_{k = \ceil{\frac{n}{r}}}^{\min(m, n)} {m \choose k} k! {n \brace k}_{\leq r}.\label{eqn:strbbprecurrence}
\end{eqnarray}

The above recurrence can be implemented using dynamic programming. The space complexity is \BigOh{m n}. The time complexity is \BigOh{m n}. The key advantage of this algorithm is that it decouples the counting of partition structures (independent of $m$, and polynomilally many) from the assignment to specific days. This dramatically reduces the state space compared to the counting based approach.

\subsection{A Recurrence Based on Direct Probability Computation}\label{sec:probrecurrence}
In this section, we develop a recurrence relation that directly computes \prob{m, n, r}, the probability that no day has more than $r$ people's birthdays incident on it in a group of $n$ people whose birthdays are distributed uniformly at random into the $m$ days of a year. Our main motivation behind this approach is to avoid counting which involves large numbers and instead derive a recurrence relation that directly computes the probability. The recurrence derived in this section is adapted from~\cite{Ramakrishna1987}.

Let \ang{m, n, r} represent a \emph{configuration} in which $n$ people's birthdays are distributed over $m$ days, and $r$ is a (free) parameter. The total number of such configurations is $m ^n$. If the distribution of birthdays is such that no day has more than $r$ birthdays incident on it, then the configuration \ang{m, n, r} is called \emph{valid}, otherwise, it is called \emph{invalid}. Let $T(m, n, r) = \card{\cset{\ang{m, n, r}}{\text{\ang{m, n, r} is valid}}}$ denote the number of valid configurations. Let $\prob{m, n, r}$ denote the probability of finding a valid configuration. Then
\begin{eqnarray}
    \prob{m, n, r} = \frac{T(m, n, r)}{m^n}.
\end{eqnarray}

For \ang{m, n, r} to be a valid configuration, it must be the case that \ang{m, n - 1, r} is a valid configuration and adding $n$-th person's birthday does not violate the valid configuration. Let $Q(m, n, r)$ denote the probability that adding $n$-th person's birthday does not violate the valid configuration, conditioned on the first $n - 1$ people's birthdays resulted in a valid configuration. Then using the formula for conditional probability, we can write
\begin{eqnarray}
    \prob{m, n, r} &=& \prob{m, n - 1, r} \cdot Q(m, n, r).\label{eqn:PQrelation}
\end{eqnarray}

We want to compute $Q(m, n, r)$. Adding $n$-th person's birthday gives an invalid configuration if and only if the day has already $r$ birthdays from the first $n - 1$ birthday distribution. Let the $n$-th person's birthday be added to the $i$-th day of the year which has exactly $r$ birthdays incident on it. Let's now compute the probability that $i$-th day has $r$ birthdays incident on it. This can happen in ${n - 1 \choose r}$ ways. The remaining $n - 1 - r$ people's birthdays must be distributed over $m - 1$ days so that no day has more than $r$ people's birthdays incident on it, which can be done in $T(m - 1, n - 1 - r, r)$ ways. Therefore the total number of ways in which $i$-th day can have exactly $r$ birthdays is ${n - 1 \choose r} \cdot T(m - 1, n - 1 - r, r)$. The number of valid configurations with $n - 1$ people with $m$ days, which defines the sample space, is $T(m, n - 1, r)$. Therefore, the probability that $i$-th day has $r$ birthdays incident on it and hence valid is
\begin{eqnarray*}
    \prob{\text{$i$-th day has $r$ birthdays incident on it} \,|\, \text{valid configuration}} &=& \frac{{n - 1 \choose r} \cdot T(m - 1, n - 1 - r, r)}{T(m, n - 1, r)},
\end{eqnarray*}
which, by symmetry, is for $1 \leq i \leq m$.

The expected number of days with $r$ birthdays incident on them is
\begin{eqnarray*}
    \Exp{\text{number of days with $r$ birthdays incident on them}} &=& m \cdot \frac{{n - 1 \choose r} \cdot T(m - 1, n - 1 - r, r)}{T(m, n - 1, r)}.
\end{eqnarray*}

Then the probability that $n$-th person's birthday in incident on a day with $r$ birthdays is given by 
\begin{eqnarray*}
    \frac{\Exp{\text{number of days with $r$ birthdays incident on them}}}{m} &=& \frac{{n - 1 \choose r} \cdot T(m - 1, n - 1 - r, r)}{T(m, n - 1, r)}.
\end{eqnarray*}
In other words, it is the probability that adding $n$-th person's birthday violates the valid configuration, conditioned on the first $n - 1$ people's birthdays resulted in a valid configuration. Therefore,
\begin{eqnarray}
    Q(m, n, r) &=& 1 - \frac{{n - 1 \choose r} \cdot T(m - 1, n - 1 - r, r)}{T(m, n - 1, r)}.
\end{eqnarray}

Recall that $\prob{m, n, r} = T(m, n, r) / m^n$. Using this in the above equation, we obtain
\begin{eqnarray}
    Q(m, n, r) &=& 1 - \frac{{n - 1 \choose r} \cdot (m - 1)^{n - 1 - r} \prob{m - 1, n - 1 - r, r}}{m^{n - 1} \prob{m, n - 1, r}}.\label{eqn:Qformula}
\end{eqnarray}

Using Equation~\ref{eqn:Qformula} in Equation~\ref{eqn:PQrelation} we obtain
\begin{eqnarray}
    \prob{m, n, r} &=& \prob{m, n - 1, r} \cdot \bracket{1 - \frac{{n - 1 \choose r} \cdot (m - 1)^{n - 1 - r} \prob{m - 1, n - 1 - r, r}}{m^{n - 1} \prob{m, n - 1, r}}} \nonumber \\
    &=& \prob{m, n - 1, r} - {n - 1 \choose r} \cdot \frac{(m - 1)^{n - 1 - r}}{m^{n - 1}} \cdot \prob{m - 1, n - 1 - r, r}.\label{eqn:Precurrence}
\end{eqnarray}

The base cases of the recurrence are:
\begin{align*}
P(m, 0, r) &= 1, \qquad \forall m \geq 1, \qquad \textup{no days, trivially valid}; \\
P(0, n, r) &= 0, \qquad \forall n > 0, \qquad \textup{no days, trivially invalid}; \\
P(1, n, r) &= \begin{cases} 1, & \text{if } n \leq r; \\ 0, & \textup{otherwise.} \end{cases} \qquad \textup{one day only}; \\
P(m, n, r) &= 0 \qquad \text{if } n > mr, \qquad \textup{impossibility by pigeonhole principle}.
\end{align*}

Taking a look at the recurrence in Equation~\ref{eqn:Precurrence}, the second term in the right hand side is a correction term that adjusts for those configurations that were valid with $n - 1$ people's birthdays but became invalid when $n$-th person's birthday is added.

The recurrence in Equation~\ref{eqn:Precurrence} can be implemented using dynamic programming. The key computation happens in evaluating the term ${n - 1 \choose r} \cdot (m - 1)^{n - 1 - r} / {m^{n - 1}}$. However, the computation can be carefully sequenced as follows so that big number operations can be avoided through simple memoization:

\begin{eqnarray}
    {n - 1 \choose r} \cdot \frac{(m - 1)^{n - 1 - r}}{{m^{n - 1}}} &=& \frac{n - 1}{n - 1 - r} \cdot \frac{m - 1}{m} \cdot \bracket{{n - 2 \choose r} \cdot \frac{\paren{m - 1}^{n - 2 - r}}{m^{n - 2}}}.\label{eqn:precomprecurrence}
\end{eqnarray}

Precomputing the values using the recurrence in Equation~\ref{eqn:precomprecurrence}, and noting that there are in total \BigOh{mn} states, the space complexity is \BigOh{mn}, and the time complexity is \BigOh{mn}, since the cost per stage is \BigOh{1}. The probability computation using this formulation is efficient both time and space complexity point of view compared to other methods we discussed previously. However, if we need to count the number of ways or desire to understand the structure of the problem, the other methods shed light in complementary ways that this method does not provide.

\section{Results and Discussion}\label{sec:results}
We implemented all the recurrences derived in the previous sections.\footnote{Our implementation is available at: https://github.com/chijulbtripathy-coder/BottleneckBirthdayProblem} We use an arbitrary precision arithmetic library so that large integer operations can be performed. The implementations all give identical results, thereby providing the validity of the theoretical derivations and implementation. The Table~1 shows the maximum number of people $n$ required for an $m$-day year so that no day of the year has more than $r \geq 1$ birthdays incident on it with probability at least 1/2, for different values of $m$ and $r$.

The recurrence in Equation~\ref{eqn:strbbprecurrence} is faster to compute both the number of ways and the probability compared to the recurrences in Equations~\ref{eqn:odtrecurrence} and~\ref{eqn:countingrecurrence}. The recurrence in Equation~\ref{eqn:Precurrence} is the most efficient one to directly compute the probability values, however, we cannot use this to compute the number of ways valid configurations can be formed.

\begin{table}[h]
  \begin{center}
   \begin{tabular}{|c|c|c|c|c|c|c|c|c|c|c|c|}
     \hline
        \diagbox[width=2.5em]{$r$}{$m$} & 10 & 25 & 50 & 100 & 200 & 365 & 500 & 1000 \\ \hline
        1 & 4 & 6 & 8 & 12 & 16 & 22 & 26 & 37 \\ \hline
        2 & 9 & 15 & 24 & 37 & 59 & 87 & 106 & 167 \\ \hline
        3 & 15 & 27 & 45 & 73 & 121 & 186 & 234 & 387 \\ \hline
        4 & 21 & 41 & 69 & 116 & 197 & 312 & 398 & 680 \\ \hline
        5 & 28 & 56 & 95 & 164 & 284 & 459 & 590 & 1030 \\ \hline
        6 & 35 & 71 & 124 & 216 & 380 & 622 & 805 & 1426 \\ \hline
        7 & 42 & 88 & 154 & 272 & 483 & 797 & 1038 & 1860 \\ \hline
        8 & 49 & 104 & 185 & 330 & 591 & 984 & 1286 & 2325 \\ \hline
        9 & 57 & 121 & 217 & 390 & 704 & 1180 & 1548 & 2818 \\ \hline
        10 & 65 & 139 & 250 & 452 & 822 & 1384 & 1820 & 3334 \\ \hline
   \end{tabular}
 \end{center}
  \caption{The maximum number of people $n$ required for an $m$-day year so that no day of the year has more than $r \geq 1$ birthdays incident on it with probability at least 1/2.}
 \label{tab:bottleneckBirthday}
\end{table}

Below we give a comprehensive comparison of our approaches. Table \ref{tab:all_approaches} compares all four approaches for the largest test case $(m, n, r) = (365, 1384, 10)$. The speedup for the direct probability recurrence is due to standard floating-point arithmetic (as opposed to arbitrary-precision arithmetic) in addition to the improved time and space complexities. Also, it is numerically stable due to the probability domain computation.

\begin{table}[h]
\centering
\caption{Sample Performance Comparison: All Four Approaches for $(m,n,r) = (365, 1384, 10)$}
\label{tab:all_approaches}
\small
\begin{tabular}{lcccc}
\toprule
Approach & Time (s) & Speedup & Time Complexity & Space Complexity \\
 & & vs. S3 & & \\
\midrule
Counting Recurrence (Section~\ref{sec:countingrecurrence}) & 855.592 & 1$\times$ & $O(m^2n^2r)$ & $O(mn^2)$ \\
Probability Recurrence (Section~\ref{sec:recOneDay}) & 202.177 & 4$\times$ & $O(mnr)$ & $O(mn)$ \\
Restricted Stirling Recurrence (Section~\ref{sec:restrstr}) & 16.336 & 52$\times$ & $O(mn)$ & $O(mn)$ \\
Direct Probability Recurrence (Section~\ref{sec:probrecurrence}) & \textbf{0.139} & \textbf{6,155$\times$} & $O(mn)$ & $O(mn)$ \\
\bottomrule
\end{tabular}
\end{table}

Based on the above analysis, the direct probability recurrence is the fastest approach. However, the restricted Stirling recurrence provides elegant combinatorial insight connecting to the partition theory, and is the second fastest approach empirically (due to the use of arbitrary-precision arithmetic), with same asymptotic time and space complexities as those for the direct probability recurrence.

\section{Conclusion}\label{sec:concl}
In the article, we provided a survey of methods as well as developed novel algorithms for doing exact computations of the number of ways valid configurations can be generated and the probability computation for the Bottleneck Birthday Problem. We also, derived a recurrence from the first principles for counting the number of ways valid configurations made, and derived another recurrence by using restricted Stirling numbers of the second kind. We discussed their time and space complexities and provided insights into the usefulness of each method. Our findings clearly suggest that the probability based method is the fastest, numerically stable and scalable one. We note that \bbp applies to numerous practical problems involving load balancing in distributed systems, hash table design, resource allocations, and other areas in science and engineering. In the future, we would like to extend our studies to non-uniform distributions of birthdays and multi-dimensional extensions (\eg, person's propensity for a certain day). Other approaches using FFT-based implementation and generating functions, as well as approximation algorithms for \bbp, can be interesting extensions.

\section*{Acknowledgments}
I am thankful to Dr.~Chittaranjan Tripathy for introducing me to discrete mathematics, birthday problem and then suggesting me that \bbp can be an interesting problem to work on. I am thankful to Dr.~Shashidhara K.~Ganjugunte for numerous valuable comments on the manuscript.

\bibliographystyle{unsrt}
\bibliography{references}  

\end{document}